\documentclass[aps, prx, preprint, floatfix, longbibliography]{revtex4-2}

\usepackage{graphicx}
\usepackage{dcolumn}
\usepackage{bm}
\usepackage{amsmath, amssymb}
\usepackage{natbib}
\usepackage[version=3]{mhchem}
\usepackage{hyperref}
\usepackage[usenames,dvipsnames]{xcolor}
\usepackage{ulem}

\definecolor{cream}{RGB}{222,217,201} 
\newcommand{\cb}{\textcolor{black}}

\begin{document}



\title{Probing wetting properties with self-propelled droplets} 

\author{Bernardo Boatini}
\email{b.boattini@gmail.com}
\thanks{Corresponding author}
\author{Cristina Gavazzoni}
\author{Leonardo Gregory Brunnet}
\author{Carolina Brito}
\affiliation{Instituto de F\'isica, Universidade Federal do Rio Grande do Sul, Postal Code 15051, CEP 91501-970, Porto Alegre, Rio Grande do Sul, Brazil}


\begin{abstract}
Wetting phenomena are relevant in several technological applications, particularly those involving hydrophobic or hydrophilic surfaces. Many substrates support multiple wetting states depending on surface conditions or droplet history —a behavior known as metastability. This feature is crucial both for its theoretical complexity and for its relevance in practical applications that rely on controlling metastable states. While several experimental and computational techniques have been developed to study metastability, they tend to be complex or computationally expensive. In this work, we introduce an alternative approach based on concepts from active matter physics. We investigate the wetting behavior of a droplet placed on a pillared surface using a \cb{3-state} Cellular Potts model with a polarity term that mimics a self-propelled droplet. Applying this model to a pillared substrate with known metastable wetting states, we demonstrate that increasing activity enables the droplet to traverse free energy barriers, explore consecutive metastable states, and eventually suppress metastability entirely. Our results show that activity reduces the disparity between dry and wet states and provides a reliable framework for identifying and quantifying metastability through contact angle measurements.
\end{abstract}


\maketitle 


\section{Introduction} 

The study of wetting phenomena is of great interest due to its wide range of technological applications involving hydrophobic and hydrophilic surfaces. Significant effort has been made to understand droplet behavior on rough solids, aiming to design substrates with controlled wetting properties~\cite{parker2001water, cheng2005lotus, liu10, Khayat_AMI2006, chen19, gava2021}. Since the foundational work of Wenzel~\cite{Wenzel1936} and Cassie-Baxter~\cite{Cassie1944}, the concept of apparent contact angle has become central, accounting for intrinsic wettability and surface heterogeneities. Some models have since emerged based on surface chemistry and topography, yet most rely on the equilibrium assumption, which means that the droplet adopts a single, stable wetting state. In reality, however, surfaces can support metastable states, where the droplet is locally trapped, leading to effects like contact angle hysteresis~\cite{Quere2008, laz19, britoJCP2023}. Metastability is common in wetting and poses both theoretical challenges and practical relevance. For example, superhydrophobicity often relies on metastable states~\cite{Quere2008}, while super-slippery surfaces aim to eliminate them \cite{wang2020comparison}. Therefore, understanding and controlling metastability is key to engineering advanced functional substrates. Experimentally and computationally, probing metastable states is difficult, as it requires extensive trials. Some methods include observing droplets from below~\cite{Sbragaglia2007} or using varied initial conditions\cite{Koishi2009, fernandes2015, patricia2023hierarchical}. Computational tools like umbrella sampling and adaptive biasing~\cite{AMI_Marion2021} have proven useful but are often complex and resource-intensive. To overcome local energy barriers, experiments often introduce initial droplet velocity or substrate vibrations~\cite{Quere2008}. Inspired by this, a promising route is continuous, nonequilibrium energy injection—such as self-propulsion. This can be achieved via engineered surfaces, coalescence, or external fields~\cite{brochard1989motions, yao2012running, lin2024emergent, han2022infinite}. More generally, active matter systems, where internal agent dynamics drive motion, provide a compelling framework for understanding this phenomenon~\cite{Joanny_Ramaswamy_2012, khoromskaia2015motility, doostmohammadi2018active, trinschek2020thin, stegemerten2022symmetry, carenza2023motility}. While some numerical and experimental studies have explored how active suspensions alter wetting on flat surfaces~\cite{coelho2023active, perez2019active, adkins2022dynamics}, their implications for the metastability paradigm in wetting physics remain largely unexplored. In this work, we apply principles and methods from active matter physics to probe the wetting properties of a substrate and characterize its metastability. \cb{We simulate the system using a 3-state Cellular Potts Model, where activity is introduced through a polarity term in the energy functional~\cite{Beatrici_Stokes}. This term induces a time-varying directional bias in the droplet’s motion, effectively modeling a self-propelled droplet with isotropic dynamics, consistent with prior experimental observations\cite{sanchez2012spontaneous, Dense_bact, kokot2022spontaneous, michelin2023self}. The model is used to investigate wetting on a pillared substrate, whose passive (equilibrium) wetting properties are well-characterized. In the absence of activity, the so called "passive" free energy landscape is characterized by the existence of several metastable states—ranging from dry to wet—depending on its geometric roughness~\cite{fernandes2015, gava2021, AMI_Marion2021}. Our key observation is that, even in the presence of activity, the system exhibits wetting properties that closely follow the passive free energy landscape. In particular, for certain substrate roughnesses, increasing activity leads the droplet to transition between distinct wetting states that correspond to those predicted by the equilibrium minima. Above a certain activity threshold, the dependence on the initial conditions disappear, and the droplet reaches the same wetting state. We emphasize that, although our system is inherently out of equilibrium, activity acts as a probe that reveals the accessibility and robustness of wetting configurations associated with the underlying passive free energy landscape. This enables us to determine whether a substrate is metastable, estimate the number of free-energy-like local minima, and measure the corresponding contact angles. }

This work is organized as follows. Section~\ref{meta} introduces the parameters of the pillared surface studied in this work and reviews its wetting properties as established in previous studies. In Section~\ref{section_simu}, we present the numerical model used, along with the initial conditions employed in the simulations to investigate metastability. Section~\ref{section_result} presents and discusses the main results, followed by Section~\ref{conclusion}, which contains our conclusions.
\section{\label{meta}Metastability of a pillared surface}

\begin{figure}
    \centering
    \includegraphics[width=0.48\textwidth]{ 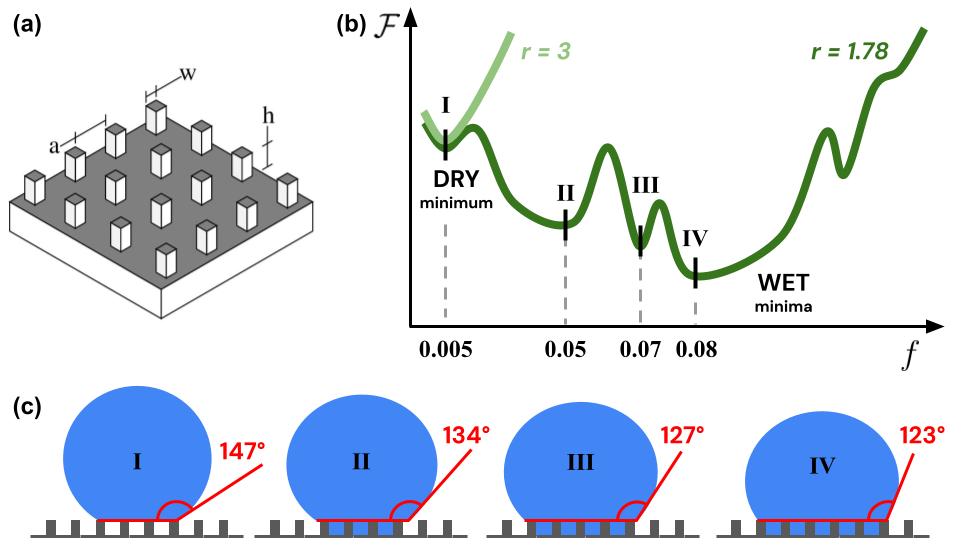}
     \caption{{\bf (a)} Definition of the geometric parameters of the pillared surface: roughness ratio $r=1+4hw/(a+w)^2$, where $a$ is the inter-pillar distance, $w$ is the pillar width and $h$ is the pillar height. {\bf (b)} Schematic representation of the free energy $\mathcal{F}$ as a function of the water fraction penetrating the substrate $f$, based on results from two distinct pillared surfaces in Ref. \cite{AMI_Marion2021}. The light-green curve, which exhibits a single minimum, corresponds to a surface with a high roughness ratio $r$, whereas the dark-green curve, displaying multiple local minima, represents substrates with lower $r$. {\bf (c)} Cross-sectional representations of the water droplet at each minimum of $\mathcal{F}$, arranged from the driest to the wettest state (I–IV) \cb{with its respective contact angle (shown in red)).}}
    \label{rev_pill} 
\end{figure}

Metastability is often understood in terms of a complex landscape of a system’s thermodynamic potential. In this framework, systems that exhibit metastability are interpreted as having local minima in their free energy landscape, separated by barriers that may depend on one or more system variables. In the context of wetting phenomena, a metastable state corresponds to a particular wetting configuration. These configurations typically fall into two categories: dry states, where the droplet does not fully infiltrate the substrate, and wet states, where the droplet penetrates the surface texture. Dry states are commonly referred to as Cassie–Baxter (CB) states, while wet states are known as Wenzel (WE) states. Each state is characterized by its apparent contact angle $\theta_C$ and a filling fraction $f$, which quantifies the portion of liquid that penetrates the substrate relative to the \cb{droplet total volume}. \cb{Pillared substrates are commonly used in both theoretical and experimental studies of wetting, as they provide a conceptually simple and tunable model that captures the essential features of roughness-dependent wetting\cite{Quere2008, Koishi2009, fernandes2015, Silvestrini2017, AMI_Marion2021}. The geometry of a pillared substrate is defined by the pillar height $h$, a pillar width $w$ and an inter-pillar distance $a$, as illustrated in Figure~\ref{rev_pill}(a). The roughness ratio—defined as the ratio of the actual surface area to the projected area—is given by $r=1+4hw/(a+w)^2$.}

Using a \cb{3-state} cellular Potts model (described in the next section) and a continuous model that minimizes the global interfacial energy, it was shown that the final wetting state of a droplet depends on its initial condition~\cite{fernandes2015} as previously observed experimentally~\cite{Quere2008, Koishi2009}. \cb{In these references, it is shown that a} droplet initially placed in a dry state would remain in a (another) dry state with a high contact angle, even though this state were not the most energetically favorable one according to the continuous model. On the other hand, if the droplet was initially placed wetting the substrate, the final state obtained in the simulation would correspond to the thermodynamically stable configuration. To better investigate this behavior, \cb{a previous study involving one of the present authors} performed a combination of constrained Monte Carlo simulations and the string method to compute the free energy profile of a liquid droplet deposited on a pillared surface as a function of the water fraction penetrating the surface $f$~\cite{AMI_Marion2021}. A schema of the result is illustrated in Figure~\ref{rev_pill}(b). The results reveal that for a certain range of geometric parameters (specifically, those associated with high surface roughness) the substrate exhibits a single free energy minimum, corresponding to a superhydrophobic wetting state. As the surface roughness decreases, the number of local minima in the free energy profile increases, indicating the emergence of one dry metastable state and multiple wet states (Figure~\ref{rev_pill}(c)).
\section{\label{section_simu}Numerical Model}

\subsection{The Cellular Potts Model (CPM)}
\begin{figure}
    \centering
    \includegraphics[width=0.45\textwidth, trim=0 0 368 0, clip]{ 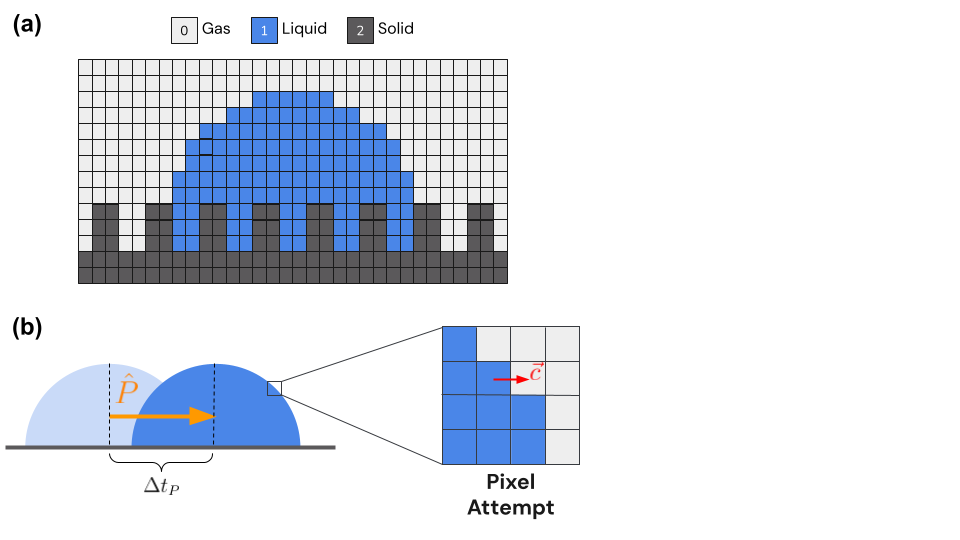}
     \caption{
     (a) Schematic cross-section of a 3D droplet over a pillared substrate in the Cellular Potts Model (CPM). Each pixel corresponds to a different state ($s_i=0, 1, 2$). (b) Representation of the activity term, Eq.~\ref{act}. The displacement of the droplet’s center of mass \cb{($\vec{P}$)}, in orange, is measured at intervals of $\Delta t_P$ (in Monte Carlo Steps, MCS). The vector \cb{$\vec{c}$(red) represents an effective displacement of a water pixel at the interface (see text for more details).}}
    \label{model} 
\end{figure}

The Cellular Potts Model (CPM), originally introduced by Graner and Glazier\cite{gra92}, has been widely used to study wetting phenomena in textured surfaces \cite{Oliveira2011,Mortazavi2013}, in particular, wetting of pillared surfaces \cite{fernandes2015, Silvestrini2017, gava2021}. We begin with the 3D CPM, defined on a simple cubic lattice. The system is governed by the following Hamiltonian:

\begin{eqnarray}
H_0 &=& \frac{1}{2} \sum_{<{\rm i},{\rm j}>} E_{s_{\rm i},s_{\rm j}}(1-\delta_{s_{\rm i},s_{\rm j}}) + \lambda \left( \sum_{\rm i} \delta_{s_{\rm i},1}-V_T \right) ^2  \nonumber \\
&+& mg \sum_{\rm i} h_{\rm i} \delta_{s_i,1}
\label{hamil}
\end{eqnarray}

\noindent where the \cb{state} $s_i \in \{0,1,2\}$ represent gas, water and solid states, respectively (Figure \ref{model}(a)).

The first term in Eq.(\ref{hamil}) represents the energy related to the presence of interfaces between sites of different types. The summation ranges over pairs of neighbors which comprise the 3D Moore neighborhood in the simple cubic lattice (26 sites, excluding the central one), \cb{$E_{s_i,s_j} = A\sigma_{s_is_j}$} is the interaction energies of sites $s_i$ and $s_j$ of different states at interfaces \cb{, $\sigma_{s_is_j}$ are the surface tensions between different phases, A is an area between them, } and $\delta_{s_i,s_j}$ is the Kronecker delta. The second term is responsible to keep the droplet volume around a target volume $V_T$. The summation is over the pixels of water and $\lambda$ represents the \cb{stiffness of the harmonic constraint, in a way that mimics water compressibility}. \cb{The last term represents the gravitational energy associated with the droplet, which, for the droplet size used in this work, is much smaller than the interfacial contributions. In practice, this term plays a role during the initial stages when the droplet is placed in the dry state, preventing the droplet from remaining suspended.}

\cb{The parameters of Eq.(\ref{hamil}) were chosen to mimic the experiments of a droplet of water on a Polydimethylsiloxane (PDMS) surface~\cite{Tsai2010}, which have $\sigma_{s_0s_1} = 70$ mN/m and $\sigma_{s_2s_0} = 25$ mN/m.} The surface tension between the solid and the liquid was obtained from Young’s relation $\sigma_{s_0,s_1} \cos(\theta_Y) = \sigma_{s_2,s_0} - \sigma_{s_2, s_1}$ with $\theta_Y=111^{\circ}$ being the contact angle on a smooth surface. The length scale is such that one lattice spacing corresponds to 1 $\mu$m and the surface tensions values are divided by 26, which is the number of neighbors that contributes to the first summation of our Hamiltonian. Therefore, $A=1\mu$m$^2$ \cb{and the interfacial energies} are given by $E_{0,1} = 2.70 \times 10^{-9}\mu$J, $E_{0,2} = 0.96 \times 10^{-9}\mu$J and $E_{1,2} = 1.93\times 10^{-9}\mu$J. The mass in a unit cube is $m^w = 10^{-15}$kg and $\lambda = 0.01 \times 10^{-9}\mu$J/($\mu$m$)^6$. This defines what we refer to as the {\it passive case droplet}, which serves as the starting point for implementing its active counterpart.
\subsection{A self-propelled droplet: the CPM with activity}

There are several ways to achieve self-propulsion in a droplet, ranging from the design of specialized surfaces to merging processes or even the application of external fields \cite{brochard1989motions, yao2012running, stamatopoulos2020droplet, lin2024emergent, han2022infinite}. In this work, we focus specifically on isotropic and Brownian-like motions, where the droplet exhibits random motion without any intentional directional control driven by external forcing. This type of behavior was already reported on experimental literature, and can be characterized as a persistent Brownian motion \cite{sanchez2012spontaneous, Dense_bact, kokot2022spontaneous}. In such systems, the energy is generally injected trough the constitutive agents of the droplet solution, which then characterizes it as a active matter suspension. To incorporate this feature in the model, an additional term is added to the variation of the Hamiltonian when a \cb{state} is changed: $\Delta H = \Delta H_0 + \Delta H_{\alpha}$;
where the additional term 
is given by \cite{Beatrici_Stokes, kabla2012collective}
\begin{eqnarray}
\Delta H_{\alpha} = -\mu\widehat{P}(t-\Delta t_P)\cdot \vec{c}
\label{act}
\end{eqnarray}
\noindent with $\mu$ representing the magnitude of an effective force corresponding to the active energy input and $\widehat{P}(t - \Delta t_P)$ \cb{is a polarization vector that tracks the normalized displacement of the droplet’s center of mass over a fixed time window $\Delta t_P = 1$ , acting as a memory term. Due to the stochastic nature of the Monte Carlo method, this vector fluctuates.} The robustness of our choice of $\Delta t_P$ is tested and shown in the Supplementary Material(SM). The vector $\vec{c}$ represents the effective displacement of a water pixel at the interface during a trial state flip. It is determined dynamically during the simulation based on the local configuration \cb{as follows. If a gas site flips into water, this site is taken as the final position, and a randomly selected neighboring water site is chosen as the initial position. Conversely, if a water site flips to gas, the initial position is the water site itself, and a neighboring water site is randomly chosen as the final position. Since only water and gas pixels are allowed to switch states, the substrate pixels remain fixed throughout the simulation. The scalar product defined in Eq.~\ref{act} corresponds to the contribution of the active force to the energy variation associated with each pixel update attempt during a Monte Carlo step (MCS).} For simplicity, we assumed that self propulsion fluctuates only on the horizontal plane \cb{parallel to the substrate}, ignoring the vertical component, \cb{orthogonal to the substrate}. The total run of a simulation is at least $2 \times 10^{5}$ MCS from which the last 10\% of the total running time are used to measure observables of interest. Each MCS is composed by $V_T$ number of trial \cb{state} flips. A \cb{state} flip is accepted with probability $ \text{min}\{1,\text{exp}(- \beta \Delta H)\} $, where $\beta = 1/T$ and $T$ acts as noise to allow the phase space to be explored. Here we used $T=13$ that was shown to have an acceptance rate of approximately $20\%$ for the passive system \cite{fernandes2015}. The simulation is exemplified in Figure \ref{model}(b).

\subsection{Pillared surface and the initial configurations}

We consider a 3D droplet on a pillared surface, with geometric parameters defined in Figure~(\ref{rev_pill})a. The pillar width is fixed at $w = 5\,\mu\text{m}$ and the height at $h = 10\,\mu\text{m}$, while the interpillar spacing $a$ is varied within the range $a \in [5\,\mu\text{m},11\mu\text{m}]$. \cb{This yields a roughness ratio ranging from $r \in [1.78, 3]$.}

\begin{figure}
    \centering    
    \includegraphics[width=0.49\textwidth, trim=0 160 0 0, clip]{ 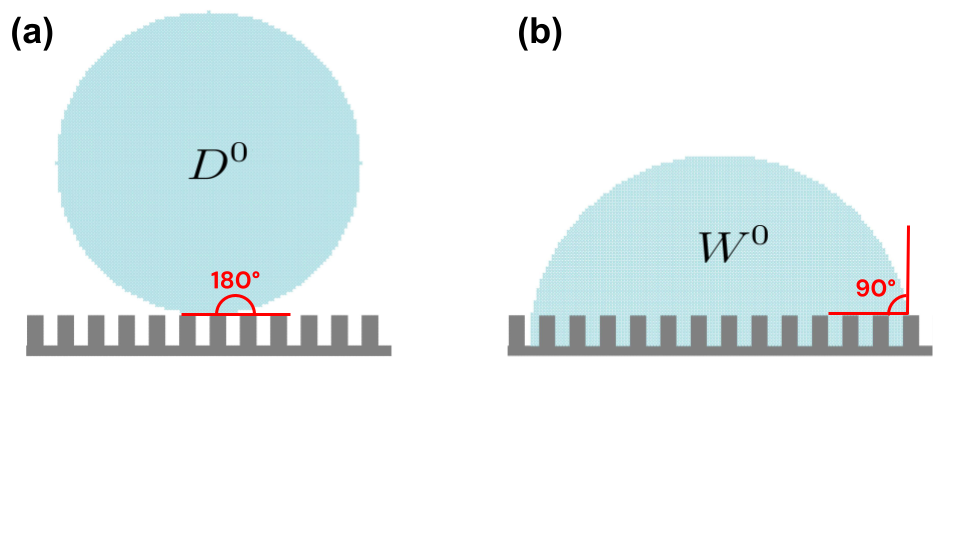}
     \caption{Two initial wetting configurations used in the Monte Carlo simulations: one in the dry state, refereed as D$^0$\cb{, with $\theta_c = 180$°} {\bf(a)}; and one in the wet state, W$^0$ \cb{where $\theta_c = 90$°} {\bf(b)}.}
     \label{CIs} 
\end{figure}

To account for the influence of initial conditions, the system is initialized in two different wetting states. The first configuration, referred to as D$^{0}$, consists of a droplet in the shape of a sphere tangentially touching the surface \cb{($\theta_c = 180^{\circ}$)}, as illustrated in Figure~(\ref{CIs})-a. The second correspond to fully wetted case, with contact angle $\theta_c = 90^{\circ}$, referred as W$^{0}$, shown in Figure~(\ref{CIs})-b. In both cases, the initial droplet volume is set to $V_T \approx V_0 = \frac{4}{3} \pi R_0^3$, with $R_0 = 50\,\mu\text{m}$. \cb{In the present study, we restrict our analysis to droplets of fixed size on various substrates, following prior work by one of us in which the free energy landscape was characterized using droplets of this same volume~\cite{AMI_Marion2021}.}

\subsection{Observables}
{\it Wetting measures}. The apparent contact angle ($\theta_c$) of the droplet is estimated by assuming a spherical shape, which is a valid approximation for the droplet size used here, where interfacial energy dominates over gravitational effects~\cite{fernandes2015, AMI_Marion2021}. Under this assumption, $\theta_c$ is defined as

\begin{equation}
    \theta_c = \arcsin \left( \frac{2H.B}{H^2 + B^2} \right),
    \label{eq:theta_c}
\end{equation}
where $B$ and $H$ are the base radius and height of the droplet, respectively, measured directly from the simulations. The validity of the spherical approximation is confirmed through circularity measurements and holds for $\mu < 8$. \cb{In particular, when $\mu > 10$ nN the droplet deviates markedly from a spherical geometry, making the definition of a single apparent contact angle $\theta_C$ ill-defined (see Section 2 of the SM).}

\cb{We emphasize that the stochastic nature of Monte Carlo simulations leads to fluctuations in the droplet’s contact angle. Consequently, even in the absence of activity, $\theta_C$ exhibits a distribution centered around a well-defined mean value, $\bar{\theta}_C$, as shown in Figure 1 of the SM. When activity is introduced, fluctuations occasionally enable the droplet to overcome a ridge and briefly explore an adjacent groove, leading to a bimodal distribution of $\theta_C$. Because this secondary state is accessed in only about 2\% of the total simulation time, we define the most frequent contact angle, denoted as $\langle \theta_C \rangle$, based on the dominant peak, and discuss this quantity in the next section. Further details on the method and its robustness are provided in the SM.}

The volume fraction $f$ of liquid penetrating the substrate is calculated as the ratio between the number of liquid pixels located below the pillar height ($V_{pu}$) and the total volume of the droplet ($V$), which fluctuates around a target volume $V_T$ during the simulation. In the steady state, this gives $f= \frac{V_{pu}}{V}$. \cb{This quantity also fluctuates and has a distribution, for the same reason as explained above for $\theta_C$. In the following we show the most frequent value of it, denoted by $\langle f \rangle$. } 

{\it Dynamical measures}. Dynamical proprieties were studied through the calculation of the mean squared displacement (MSD), defined as:

\begin{equation}
    |\vec{\Delta r} (\Delta t)|^2 = |\vec{r}(t_0 + \Delta t) - \vec{r}(t_0)|^2;
\end{equation}
where $\vec{r}(t)$ represents the position vector of the center of mass of the droplet at time $t$. The average is computed over multiple measurements of $|\Delta \vec{r} (\Delta t)|^2$, leading to $MSD = \langle |\Delta \vec{r}(\Delta t)|^2 \rangle$.
\section{\label{section_result}Results and Discussion}

\subsection{Wetting Measures}

\begin{figure*}
    \centering    
    \includegraphics[width=0.75\textwidth]{ 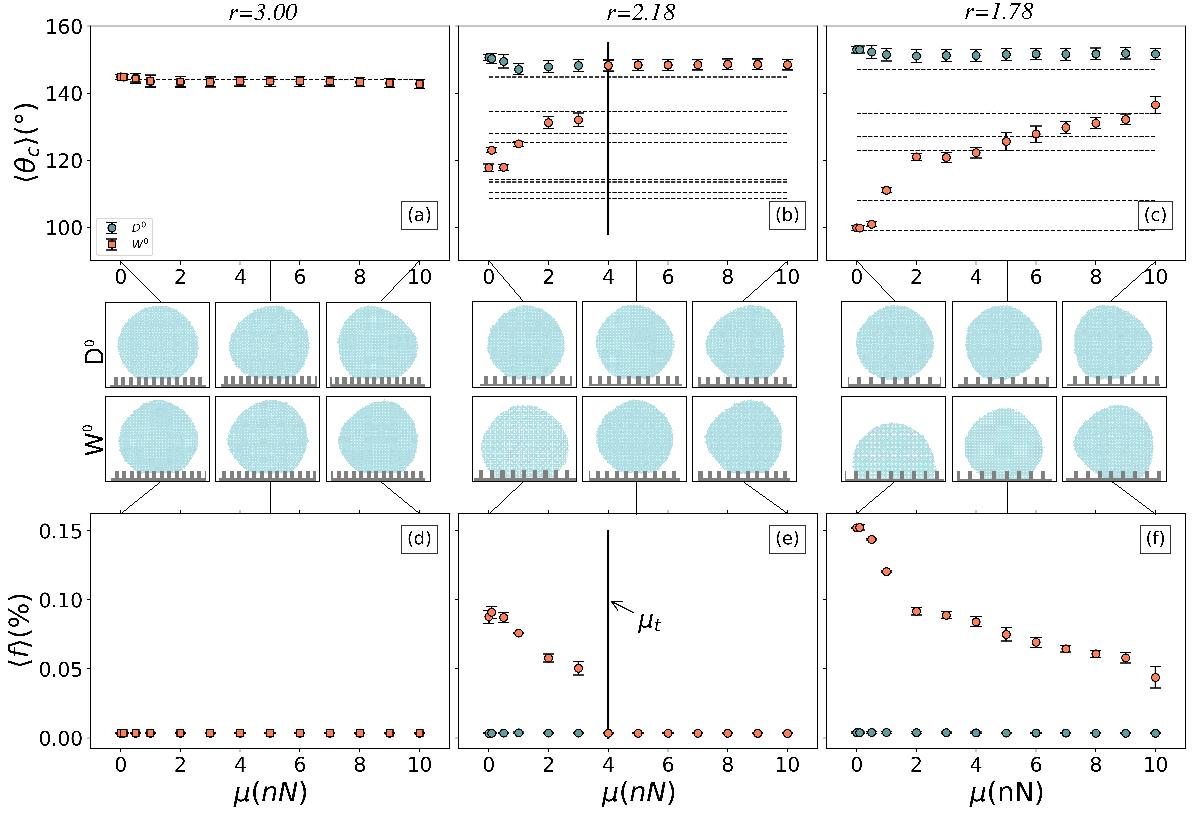}%
    \caption{\textbf{Wetting properties for three roughness values $r$.} Top row: \cb{most frequent contact angle $\langle \theta_c \rangle$} as a function of activity $\mu$. \cb{Error bars represent the standard deviation around the mean value $\langle \theta_C \rangle$.} Middle row: cross-sections of the droplet for different values of $\mu$. Bottom row: 
    Fraction of water penetrating the substrate, $\langle f \rangle$, as a function of $\mu$\cb{, with error bars representing the standard deviation around $\langle f \rangle$.} All plots are shown for two initial conditions: D$^0$ and W$^0$. The differences in $\langle \theta_C \rangle$ and $\langle f \rangle$ between these initial conditions indicate that substrates with $r=2.18$ and $r=1.78$ exhibit metastability, whereas $r=3$ corresponds to a surface with a single energy minimum. Horizontal dotted lines indicate the values of $\theta_C$ associated with the free energy minima reported in Ref.~\cite{AMI_Marion2021}.
    }
    \label{summary_wetting}
\end{figure*}

Figure~(\ref{summary_wetting}) summarizes the wetting behavior of the pillared substrates for three different roughness values $r$. The top row of the figure shows the droplet’s most frequent contact angle \cb{$\langle \theta_C \rangle$} for two initial configurations, D$^0$ and W$^0$. The middle row displays cross-sections of the droplet for different values of activity $\mu$, while the bottom row presents the fraction of water that penetrates the substrate\cb{, $\langle f \rangle$}. The first column in Figure~(\ref{summary_wetting}) corresponds to the substrate with the highest roughness ratio explored in this study, $r = 3$. The results show that: i) There is no significant difference in either \cb{$\langle \theta_C \rangle$} or the penetration fraction \cb{$\langle f \rangle$}, regardless of the initial configuration, indicating that this substrate does not exhibit metastability. ii) Similarly, \cb{$\langle \theta_C \rangle$} and \cb{$\langle f \rangle$} remain constant across all values of the activity parameter $\mu$. The cross-sectional views visually confirm that the droplet remains in a dry state, regardless of the initial condition or the value of $\mu$. Additionally, for $\mu>8$, the droplet deviates from a spherical shape, showing visible deformation, which is quantified in the SM. \cb{The} horizontal dotted line in Figure~(\ref{summary_wetting})a marks the value of the contact angle, $\theta_C$, which is the unique stable state reported in reference~\cite{AMI_Marion2021}. As briefly discussed in the previous section, that study computed the free energy profile for the pillared surface \cb{in the case without activity} and found that, for this particular roughness $r$, only a single free energy minimum exists. This minimum corresponds to a dry state and exhibits the same contact angle $\theta_C$ observed in our simulations. As the roughness ratio $r$ decreases—shown in the second and third columns of Figure~(\ref{summary_wetting})—a noticeable difference in both \cb{most frequent} contact angle \cb{$\langle \theta_C \rangle$} and penetration fraction \cb{$\langle f \rangle$} is observed depending on the initial condition. When the droplet starts from the D$^0$ configuration, the \cb{most frequent} contact angle remains high (\cb{$\langle \theta_C \rangle \approx 145^\circ$ }), and \cb{$\langle f \rangle$} $ \approx 0$ for all values of $\mu$, indicating that the droplet stays in a dry state. However, when the droplet is initialized in a wetting configuration such as W$^0$, the behavior changes. Focusing on the middle column, which corresponds to an intermediate roughness value ($r = 2.18$), the \cb{most frequent} contact angle starts at approximately \cb{$\langle \theta_C \rangle \approx 120^\circ$ } for $\mu = 0$, with \cb{$\langle f \rangle$} $\approx 10\%$. The corresponding cross section confirms that the droplet wets the substrate. As $\mu$ increases, \cb{$\langle \theta_C \rangle $} also increases, while the penetration fraction \cb{$\langle f \rangle$} decreases. At a given value $\mu_t$, the final state becomes identical for both initial conditions. This indicates that above a threshold activity $\mu_t$, the self-propulsion is sufficient to drive the droplet into the dry state, effectively overcoming the free energy barrier between metastable states. Importantly, $\mu_t$ depends on the geometric properties of the substrate.
\begin{figure}
    \centering    
    \includegraphics[width=0.98\columnwidth]{ 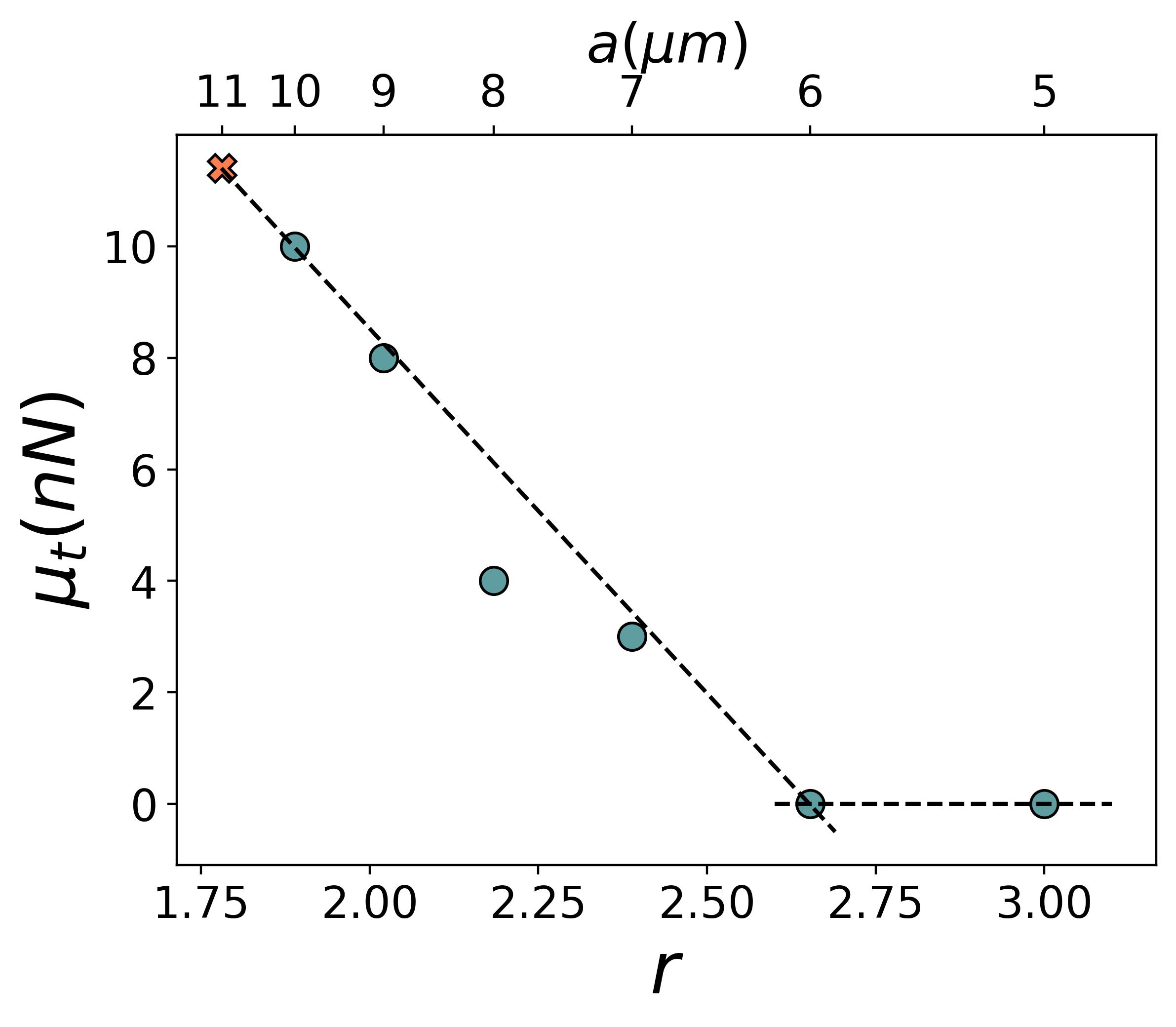}
    \caption{Threshold value $\mu_t$ is shown as a function of the substrate’s geometric parameters. \cb{Horizontal} (bottom) represents the \cb{roughness ratio $r$}, while the top axis indicates \cb{corespondent interpillar distance $a$}. When $\mu_t = 0$, the substrate exhibits a single energy minimum. For $\mu_t > 0$, multiple minima emerge, signaling \cb{a} transition from a non-metastable substrate to one that supports metastable states. \cb{Circles indicate actual data points, while the orange cross represents a prediction based on a linear fit given by $\mu_t=-13.1~r+34.7$ and shown as a dashed line.}}
    \label{muT}
\end{figure}

For the smallest roughness ratio studied ($r = 1.78$) \cb{and the range of $\mu$ values studied here,} the droplet always remains in a wet state when initialized in W$^0$ although it transitions through a series of distinct wetting states characterized by varying values of \cb{$\langle \theta_C \rangle $} and \cb{$\langle f \rangle$}, as shown in Figure~(\ref{summary_wetting})-c,f. \cb{As $\mu$ increases, the wetting states shift toward the drier regime, which is visually evident in the droplet cross sections that display a gradual reduction in interpillar filling.}

We note that these figures also include several horizontal dotted lines, which represent the contact angles of metastable states identified in the reference~\cite{AMI_Marion2021}. \cb{We recall that these metastable states are defined with respect to the equilibrium scenario, in which the free energy is calculated without the influence of activity~\cite{AMI_Marion2021}.} This suggests that on substrates with multiple free energy minima the self-propelled droplet is able to \cb{reveal the} range of metastable states as $\mu$ increases.

In the SM, we examine the robustness of the \cb{visited} metastable states by initializing the droplet in an alternative wetting configuration\cb{, for which the droplet penetrates more the substrate}. The results indicate that the same metastable states are explored as in the case starting from the \cb{initial} configuration W$^0$. This suggests that the self-propelled droplet is capable of consistently probing \cb{a large set of wetting} states. However, some of the metastable states reported in reference~\cite{AMI_Marion2021} are not accessed in our simulations.

Figure~(\ref{muT}) shows the relationship between the threshold activity $\mu_t$ and the roughness ratio $r$ for various substrate geometries. Parameter $\mu_t$ serves as an indicator of metastability: $\mu_t = 0$ signifies the absence of metastability in the substrate, while $\mu_t > 0$ indicates the existence of metastable states. Moreover, the results reveal that $\mu_t$ decreases approximately linearly as $r$ increases, suggesting that greater self-propulsion is required for the droplet to escape from deeper free energy wells associated with lower roughness. This also implies that, for sufficiently high activity, the droplet can overcome energy barriers and effectively suppress metastability, reaching the dry state. We repeated the measurements of $\langle \theta_C \rangle$ \textit{vs.} $\mu$ for larger values of $\Delta t_P=10, 100$, and show in the Supplementary Material that the results remain highly robust. This indicates that the self-propelled droplets are effectively probing the substrate’s free energy landscape \cb{in the absence of activity,} rather than simply undergoing random, activated motion.
\begin{figure*}
    \centering
    \includegraphics[width=0.80\linewidth]{ 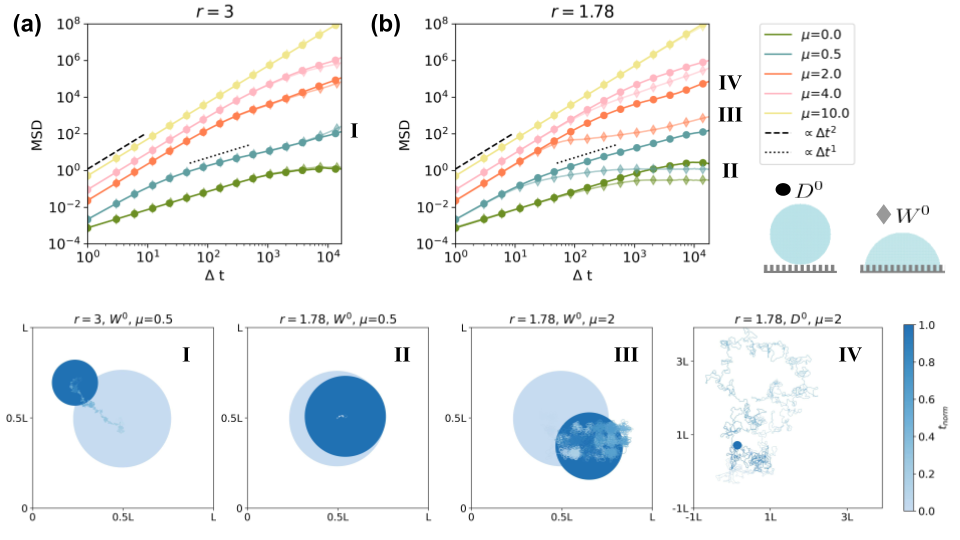}
    \caption{\textbf{Dynamical proprieties.} Top row: MSD {\it vs} $\Delta t$ multiple $\mu$(colors) curves for {\bf (a)} $r=3$, {\bf (b)} $r=1.78$. Droplets were initialized in 2 different initial wetting conditions: D$^0$ (circle marker) and W$^0$ (diamond marker). Bottom row: Representative trajectories associated with four MSD curves (I, II, III and IV) of droplets starting \cb{at} $t=0$ (light blue) until $t=t_{max}$ (dark blue). The circles reproduces the initial and final droplet circumference over the pillars. The parameter sets are: {\bf (I)} $r=3$, $W^0$,$\mu=0.5$; {\bf (II)} $r=1.78$,$W^0$,$\mu=0.5$; {\bf (III)} $r=1.78$,$W^0$,$\mu=2$; (IV) $r=1.78$,$D^0$,$\mu=2$}
    \label{MSD} 
    \end{figure*}

\subsection{Dynamical Measures}

The mean squared displacement (MSD) curves of the self-propelled droplet are shown in Figure~\ref{MSD}. Persistent Brownian motion typically exhibits two characteristic regimes: an initial ballistic regime ($\text{MSD} \propto \Delta t^2$) followed by a diffusive regime ($\text{MSD} \propto \Delta t^1$), though intermediate behaviors may also arise. The transition from ballistic to diffusive motion reveals the persistence length $l_P$ of the self-propelled droplet, representing the typical distance it travels before its polarization direction changes. For the substrate with roughness ratio $r=3$ (Figure~\ref{MSD}a), no significant differences in droplet dynamics are observed between the two initializations, D$^0$ and W$^0$. This supports the notion that this substrate does not exhibit metastability, as discussed in the previous section. In contrast, for the case $r=1.78$, shown in Figure~\ref{MSD}b, notable behaviors emerge at relatively low values of activity $\mu$. At short times, the ballistic regime appears to be independent of the initial condition. However, beyond this initial phase, the MSD curves diverge, with one trajectory exhibiting higher motility than the other, corresponding to D$^0$ and W$^0$, respectively. This effect is particularly evident for $\mu < 4$ in Figure~\ref{MSD}b. \cb{As $\mu$ increases, the MSD curves converge, indicating that this observable is not a reliable probe of the underlying metastability—unlike the contact angle and the volume fraction, which effectively capture signatures of metastability, as shown in Figure~(\ref{summary_wetting}).}

Across all cases in Figure~\ref{MSD}, increasing $\mu$ extends the duration of the ballistic regime, consistent with \cb{the model construction and in agreement with} observations in the literature \cite{Beatrici_Stokes, kabla2012collective}. Furthermore, the regime following ballistic motion varies depending on the combination of roughness $r$, activity $\mu$, and the initial wetting condition. These variations are qualitatively illustrated by representative droplet trajectories in Figures~\ref{MSD}I-IV. For $r=3$—a non-metastable substrate all values of $\mu$ eventually lead the droplets into a diffusive regime, independent of the initial condition\cb{: as both D$^0$ and W$^0$ overlap, the MSD is not able to distinguish between initialization}. In this scenario, $\mu$ primarily determines the scale of \cb{the} persistent motion\cb{, but leading to the same apparent diffusion}, as exemplified in Figure~\ref{MSD}I. Conversely, for the metastable substrate with $r=1.78$, the initial wetting condition plays a critical role. The W$^0$ initialization can lead to either a trapped state\cb{, where the droplet get stuck in the position} (Figure~\ref{MSD}II); or a caged state\cb{, in which the droplet alternate between trapped trajectories and persistent ones} (Figure~\ref{MSD}III), depending on the value of $\mu$. Meanwhile, droplets initialized with D$^0$ display consistent motile behavior across both substrate types, resulting in a typical diffusive pattern (Figure~\ref{MSD}IV)\cb{; which distinguish from the caged one since there is no sub-diffusive regime after the ballistic one (Figure~\ref{MSD}b)}. We tested the robustness of the dynamical behavior with respect to the memory parameter $\Delta t_P$ (see SM) and found that the MSD is highly sensitive to its value. Larger $\Delta t_P$ enhances the polarization term\cb{, given by the model construction itself}, resulting in a delayed transition to the diffusive regime. This sensitivity suggests that MSD alone cannot reliably infer the metastable properties of the substrate\cb{, since higher $\Delta t_P$ affect the distinction between D$^0$ and W$^0$ MSD curves, which not happen with the wetting measures for the same inicializations}. We argue that $\Delta t_P$ should be chosen such that the persistence length $l_P$ is compatible with the substrate's geometric scale. For instance, if $l_P \gg a$, meaningful information about substrate-induced metastability cannot be extracted from the MSD. Nevertheless, regimes characterized by highly linear droplet motion—where $l_P$ is large—may remain physically relevant in other contexts, such as those involving nematic active systems~\cite{coelho2023active, stegemerten2022symmetry}.
\section{\label{conclusion}Conclusions}

In this work, we present a novel approach to investigating wetting metastability by incorporating active matter dynamics into a cellular Potts framework~\cite{Beatrici_Stokes, kabla2012collective}. \cb{Using simulations of self-propelled droplets driven by an effective polarization model~\cite{Beatrici_Stokes, kabla2012collective}, we show that activity enables the droplets to effectively probe the substrate’s free energy landscape—computed in the absence of activity—rather than merely undergoing random, activated motion.}

We quantified the wetting behavior of the substrate by measuring the droplet’s \cb{most frequent} contact angle $\langle \theta_C \rangle$ and the liquid penetration fraction $\langle f \rangle$, while the mean squared displacement (MSD) was used to assess the droplet’s dynamics. \cb{Substrates with smaller roughness tested on this work} exhibits configuration-dependent wetting states, which is a hallmark of metastability. Notably, beyond a critical activity threshold, $\mu_t$, droplets converge to a single dry state, irrespective of initial conditions. This transition implies that self-propulsion energizes droplets sufficiently to surmount inter-state barriers. The threshold $\mu_t$ itself emerges as a quantitative descriptor of substrate metastability, scaling linearly with geometric parameters like the roughness ratio, $r$. \cb{A limitation of our study is the use of a simplified activity term to mimic activity, which assumes that an internal active agent propels the droplet by moving its center of mass. It would be valuable to explore a microscopic model of active matter~\cite{carenza2023motility} to better understand how variations in active particle density and their corresponding phases influence droplet motion and its macroscopic behavior. It would also be valuable to investigate the influence of droplet volume on how activity probes the substrate’s metastability landscape. While a larger, more massive droplet may require a higher level of activity to overcome free energy barriers between metastable states, we expect that the fundamental mechanism by which activity explores the landscape would still apply. Testing the extent and limits of this hypothesis would be an interesting direction for future work.}

\cb{An important next step would be to test our approach on more complex substrates to determine whether our findings extend to systems with other interesting wetting behaviors, such as oleophobic surfaces~\cite{tuteja2007designing, liu2014turning}. Finally, experimental tests of these predictions, for example using active droplets driven by bacteria~\cite{Dense_bact}, would be essential to validate the model and evaluate its applicability to real-world systems.}

\section*{Data availability}
The data that supports the findings of this study are available within the article and its supplementary material.
\section*{Acknowledgements}
We thank C. Beatrici for stimulating discussions about this work. We thank the Brazilian agencies Coordenação de Aperfeiçoamento de Pessoal de Nível Superior (CAPES) and the Conselho Nacional de Desenvolvimento Científico e Tecnológico (CNPq) for their support. CB and LB acknowledge funding under grant number CNPq-443517/2023-1. We also thank the supercomputing laboratory \href{https://pnipe.mcti.gov.br/laboratory/19775}{VD Lab} at IF-UFRGS and the \href{https://sites.google.com/nyu.edu/nyu-hpc/home?authuser=0}{NYU IT High Performance Computing} resources, services, and staff expertise, for computer time.

\bibliography{rsc} 

\end{document}